\documentclass[showpacs,aps,prd,preprint,nofootinbib,showkeys,unsortedaddress,raggedbottom]{revtex4-1}
\pdfoutput=1

\usepackage{bm}
\usepackage{amsmath}
\usepackage{graphicx}
\graphicspath{{figures/}}
\usepackage{subfig}
\usepackage{float}
\usepackage[usenames,dvipsnames]{color}
\definecolor{darkblue}{RGB}{0,0,196}
\usepackage[colorlinks=true,linkcolor=darkblue,citecolor=darkblue,urlcolor=darkblue]{hyperref}

\usepackage{setspace}

\usepackage{footmisc}

\usepackage[makeroom]{cancel}

\usepackage[utf8]{inputenc}

\def\be{\begin{equation}}
\def\ee{\end{equation}}
\def\ba{\begin{eqnarray}}
\def\ea{\end{eqnarray}}

\usepackage{lineno}

\begin{document}

\title{Photon production and elliptic flow from momentum-anisotropic quark-gluon plasma}

\author{Babak S. Kasmaei and Michael Strickland}

\affiliation{Department of Physics, Kent State University, Kent, OH 44242 United States}

\begin{abstract}
The emission of real photons from a momentum-anisotropic quark-gluon plasma (QGP) is affected by both the collective flow of the radiating medium and the modification of local rest frame emission rate due to the anisotropic momentum distribution of partonic degrees of freedom. In this paper, we first calculate the photon production rate from an ellipsoidally momentum-anisotropic QGP including hard contributions from Compton scattering and quark pair annihilation and soft contribution calculated using the hard thermal loop (HTL) approximation. We introduce a parametrization of the nonequilibrium rate in order to facilitate its further application in yield and flow calculations. We convolve the anisotropic photon rate with the space-time evolution of QGP provided by 3+1d anisotropic hydrodynamics (aHydro) to obtain the yield and the elliptic flow coefficient $v_2$ of photons from QGP generated at Pb-Pb collisions at LHC at 2.76 TeV and Au-Au collisions at RHIC at 200 GeV. We investigate the effects of various parameters on the results. In particular we analyze the sensitivity of results to initial momentum anisotropy. 
\noindent

\end{abstract}

\maketitle

\section{Introduction}

One of the challenges in developing a more complete picture of the dynamics of the many-body system generated at heavy-ion collision experiments is the lack of reliable information about initial conditions of the system due to the strong interactions among the degrees of freedom, relevance of many effects at extreme conditions, and the loss of information in a pseudo-thermalized system. The idea that internally generated electromagnetic probes of the strongly interacting matter can provide less distorted, or at least extra, information about the phases and evolution dynamics of the system has been suggested since the introduction of the notion of quark-gluon plasma \cite{Feinberg:1976ua,Shuryak:1978ij} and has been developed through decades by many researchers. In particular, real photons produced in heavy-ion collisions can be emitted from different sources and stages and are affected by various characteristics of the QCD matter \cite{Oliva:2017pri, Iatrakis:2016ugz, David:2019wpt, Gale:2018ofa, Schafer:2019edr, Hauksson:2017udm, Shen:2014nfa, Turbide:2003si, Dusling:2009ej, Linnyk:2015tha, Basar:2012bp, Muller:2013ila, PhysRevC.84.054906, Kapusta:1991qp, Nadeau:1992cn, Heffernan:2014mla, Khachatryan:2018ori, Monnai:2015qha, Monnai:2014kqa, Benic:2016uku, Fukushima:2012fg, JalilianMarian:2005zw, Berges:2017eom, Fries:2005zh, Steffen:2001pv, vanHees:2014ida, McLerran:2015mda, Ayala:2017vex, Fries:2002kt, Ghiglieri:2013gia, Neumann:1994xn, Srivastava:1999ct, Dumitru:1994vc, Baier:1997xc, Arnold:2001ms, Hung:1996mq, Singh:2015hqa, Tuchin:2014pka, Ghiglieri:2016tvj, McLerran:1984ay, STRICKLAND1994245, Sakaguchi:2014ewa, Hidaka:2015ima, Basar:2014swa, Turbide:2005bz, Baym:2017qxy, Ruan:2014kia, Schenke:2006yp}. In recent years, phenomenological studies of photon emission using hydrodynamic modeling of heavy-ion collisions have been developing \cite{Chatterjee:2005de, Chatterjee:2013naa, Dasgupta:2017fns,Bhattacharya:2015ada, Paquet:2015lta, Shen:2013cca, Dion:2011pp, Kim:2016ylr, Holopainen:2011pd, Vujanovic:2016anq, Gale:2014dfa} towards connecting theoretical ideas to experimental data. However, simultaneous quantitative description of photon yield and flow coefficients has been challenging \cite{Gale:2018ofa}. Current levels of uncertainty in experimental data of real photons at RHIC \cite{Adare:2018wgc, Khachatryan:2018evz, Adare:2015lcd} and LHC \cite{Acharya:2018bdy} also limits the extent of reliable interpretations of theoretical predictions. 

One important feature of nonequilibrium QGP, as suggested by different microscopic models and successful phenomenological studies, is the anisotropy of the local rest frame (LRF) momentum distributions for partonic degrees of freedom \cite{Strickland:2014pga}. Development of relativistic anisotropic hydrodynamics \cite{Florkowski:2010cf, Martinez:2010sc, Alqahtani:2017tnq, Alqahtani:2017jwl, Alqahtani:2017mhy, Almaalol:2018gjh} has allowed for consistent incorporation of momentum anisotropy in the collective dynamics of QGP. Modification of the photon production rate due to momentum anisotropy has also been studied \cite{Schenke:2006yp} using the parametrization introduced in \cite{Romatschke:2003ms} where momentum anisotropy is described by spheroidal deformation of conventional isotropic distributions. Using the photon rate and an early version of anisotropic hydrodynamics using a spheroidal parametrization, the yield and $v_2$ of photons from QGP has been studied previously \cite{Bhattacharya:2015ada}. In this paper, for first time we include in LRF photon rate calculation the momentum anisotropy in transverse direction in the form of ellipsoidal deformation of isotropic distributions. This is done using an efficient method, introduced in our previous paper \cite{Kasmaei:2018yrr}, to calculate the quark self-energies in anisotropic medium. In this paper we further introduce a suitable ansatz to encode the numerically calculated nonequilibrium photon rate values in a compact form which facilitates its use for hydrodynamic computation of photon yield and flow. We convolve the LRF photon rate  with the state of art 3+1d relativistic anisotropic hydrodynamic model with a quaisparticle equation of state (EOS) \cite{Alqahtani:2017mhy} and obtain results for photon yield and $v_2$. We report the results for different centrality classes, initial momentum anisotropies, Pb-Pb collisions at LHC and Au-Au collision at RHIC. We also investigate the uncertainties due to our approximations in using aHydro output for photon rate calculations.

\section{Photon production rate}

In the local rest frame of a QGP fluid element, we consider the production rate of real photons from the hard processes of Compton scattering $qg \rightarrow q\gamma$ and pair annihilation $q\bar{q} \rightarrow g\gamma$ calculated at tree level and from soft processes calculated within HTL perturbation theory \cite{Braaten:1989mz} at leading order. The photon production rate is then the combination of hard and soft contributions separated at a momentum scale $p^*$ which serves as IR/UV cutoff for hard/soft processes. 
With nonequilibrium momentum distributions $f_{q/g}({\bf k})$ for partonic degrees of freedom, the production rate of photons with momentum $q$ from Compton scattering is  
\ba
q\frac{dR^{\gamma}_{\rm{Com}}}{d^3q} = -128\pi^3\alpha_s\alpha_{\rm em} \sum_{j\in \{u,d\}}e^2_j \int_{\bf k_1}\frac{f_q({\bf k_1})}{k_1} \int_{\bf k_2}\frac{f_g({\bf k_2})}{k_2} \int_{\bf k_3}\frac{1-f_q({\bf k_3})}{k_3} \\
\times\delta^4\left( K_1 - K_2 - K_3 - Q \right) \left[ \frac{s}{t}+ \frac{t}{s} \right] \nonumber , 
\ea
and the rate from annihilation process is calculated as
\ba 
q\frac{dR^{\gamma}_{\rm{Ann}}}{d^3q} = 64\pi^3\alpha_s\alpha_{\rm em} \sum_{j\in \{u,d\}}e^2_j \int_{\bf k_1}\frac{f_q({\bf k_1})}{k_1} \int_{\bf k_2}\frac{f_q({\bf k_2})}{k_2} \int_{\bf k_3}\frac{1+f_g({\bf k_3})}{k_3} \\
\times\delta^4\left( K_1 - K_2 - K_3 - Q \right) \left[ \frac{u}{t}+ \frac{t}{u} \right], \nonumber
\ea
where $s, t$, and $u$ are usual Mandelstam variables. In this paper we use $\alpha_s = 0.3$ and $\alpha_{\rm em} = 1/137$. 
The IR cutoff $p^*$ is imposed on the momentum transfer $P = K_1 - Q$ of hard Compton and annihilation processes.

The soft contribution to the photon rate is given by \cite{Baier:1997xc}
\be 
q\frac{dR^{\gamma}_{\rm{Soft}}}{d^3q} = \frac{i}{2(2\pi)^3}{\rm Tr}\Pi_{12}(Q) ,
\ee
where the hard loop result for the trace of (12) element of the photon polarization tensor is
\be 
i{\rm Tr}\Pi_{12}(Q) = - \sum_{j\in \{u,d\}}8 e^2 e^2_j N_c \frac{f_q({\bf q})}{q} \int_{\bf p}^{p^*} Q_{\nu}\tilde{W}^{\nu}({\bf p}), 
\ee
where $p^*$ acts as the UV cutoff for the integration and we have defined \cite{Schenke:2006yp} 
\ba  
\tilde{W}^{\nu}({\bf p}) &=& \left[ {W^{\nu\alpha}}_\alpha (P) - {W_\alpha}^{\nu\alpha}(P) + {W_\alpha}^{\alpha\nu}(P)   \right]_{p_0=p({\bf \hat{p}.\hat{q}})} \\ 
W_{\alpha\beta\gamma} &=& \frac{P_\alpha - \Sigma_\alpha(P)}{(P- \Sigma(P))^2} {\rm Im}\left[\Sigma_\beta(P) \right] \frac{P_\gamma - \Sigma^*_\gamma(P)}{(P- \Sigma^*(P))^2},
\ea
and the quark self-energy is given by
\be 
\Sigma(P) = \frac{C_F}{4} g_s^2 \int_{\bf  k} \frac{\hat{f}({\bf k})}{|{\bf k}|}\frac{K.\gamma}{K.P},    \label{qse}
\ee
in which the combined distribution $\hat{f}({\bf k}) = 4 f_{\rm g}({\bf k}) + 2 \left( f_{\rm q} \left( {\bf k}\right) + f_{\rm \bar{q}} \left( {\bf k} \right) \right) $ is used \cite{Mrowczynski:2000ed}.

The photon rate for anisotropic momentum distributions with spheroidal parametrization has been calculated previously \cite{Schenke:2006yp}. In this paper we extend the results to include an ellipsoidally anisotropic distribution in the LRF 
\be 
f({\bf k}) = f_{\rm iso}\left(  \frac{k}{\Lambda}\sqrt{1+ \xi_1 ({\bf \hat{k}.\hat{n}_1})^2  + \xi_2 ({\bf \hat{k}.\hat{n}_2})^2} \right),
\ee
with which the photon rate acquires dependence on two more variables $\xi_2$ and $\phi_q$. $\Lambda$ is the temperature-like scale. In some equations or plots we represent momenta as scaled by $\Lambda$ i.e. $\hat{q} = q/\Lambda$. For $f_{\rm iso}$, Fermi-Dirac/Bose-Einstein distribution is used for quarks/gluons.

The generalization of hard contributions to the ellipsoidal case is straightforward. On the contrary, calculation of the quark self-energy \eqref{qse} for an ellipsoidal anisotropy was shown \cite{Kasmaei:2016apv} to be more tedious than the spheroidal case \cite{Schenke:2006fz}. In a previous paper \citep{Kasmaei:2018yrr} we introduced an efficient method to calculate the integral \eqref{qse} for general forms of anisotropic momentum distributions which makes it possible to obtain the results for photon rates in this paper. 

\begin{figure}%
    \centering
    \subfloat[]{{\includegraphics[width=0.45\linewidth]{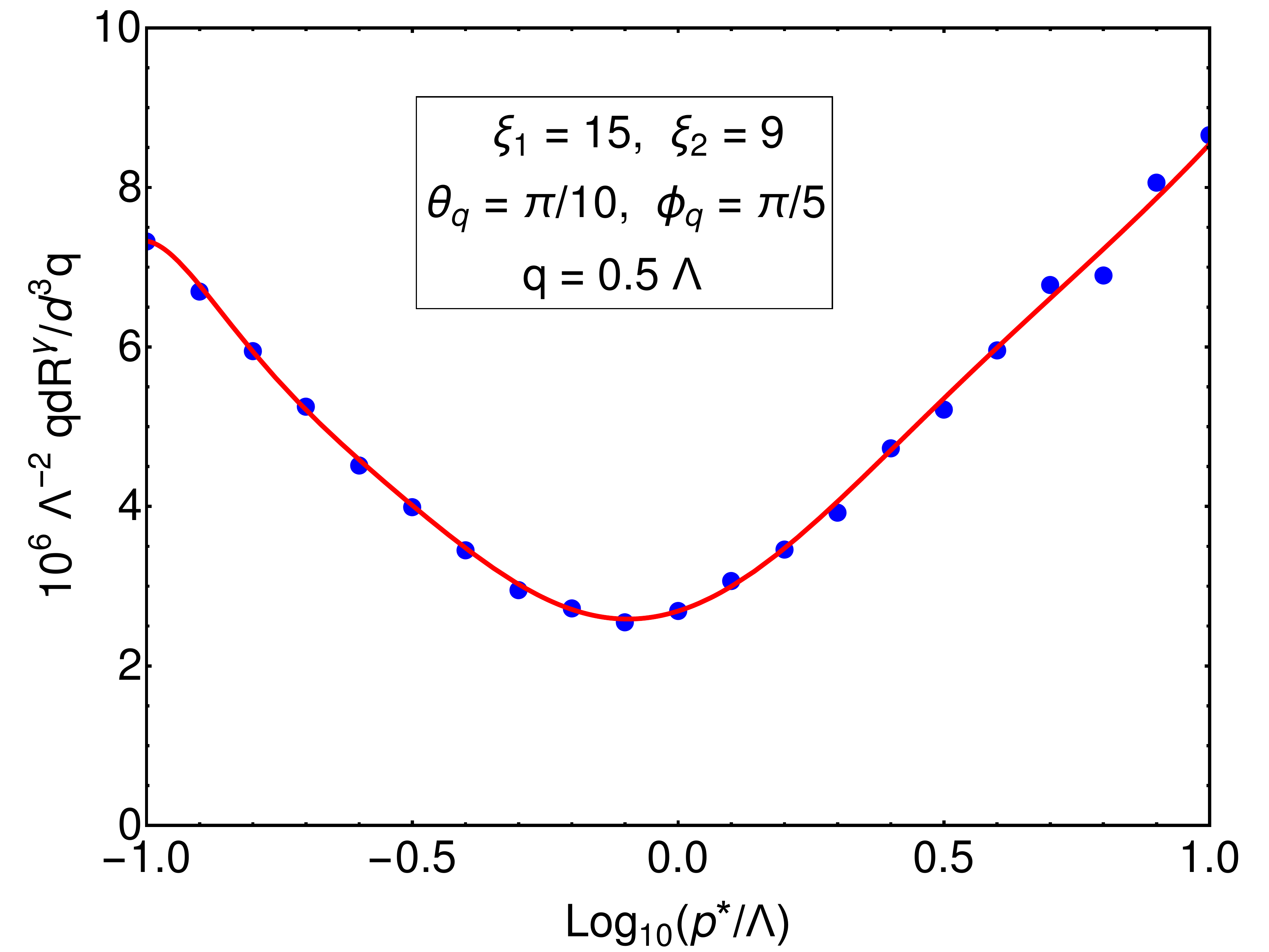} }} \label{plot:pstarplt}
    \subfloat[]{{\includegraphics[width=0.45\linewidth]{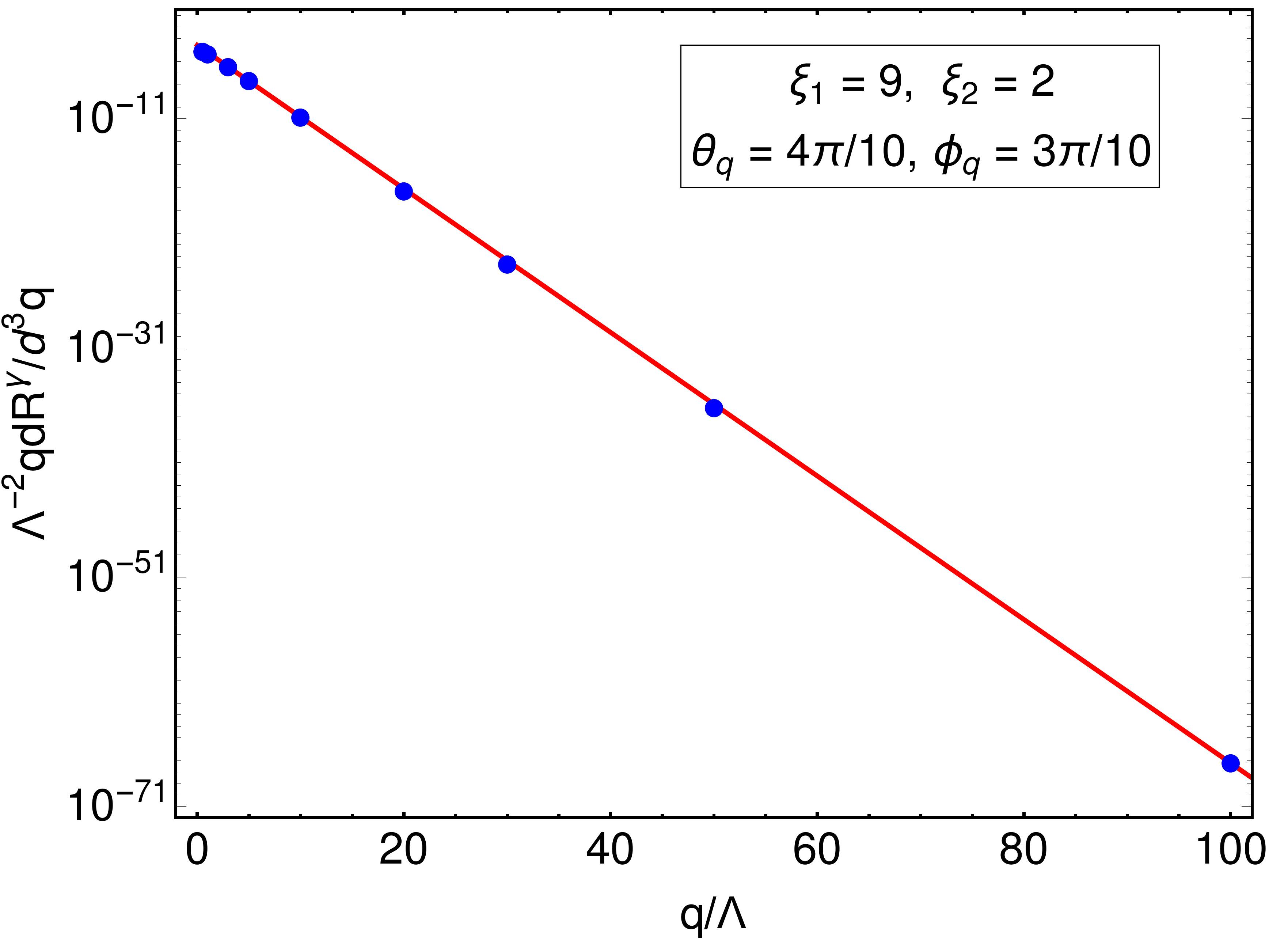} }} \label{plot:qansatz}
    \caption{(a) Example case for total photon rate as a function of hard/soft separation momentum $p^*$. Points represent numerically calculated results and solid curve is a fitted polynomial to the points. (b) An example case for fitting exponential ansatz \eqref{qansatz} to numerically calculated values of photon rate.}%
    \label{plot:fig1a2}%
\end{figure}

The total photon rate calculated with separation of hard and soft momenta (Braaten-Yuan method \citep{Braaten:1991dd}) depends on the cutoff $p^*$. We select the point for which the total rate as a function of $p^*$ has its minimum. This idea is shown in the left panel of Fig.~\ref{plot:fig1a2} for an example set of parameters where we fit a polynomial to the numerically calculated points and then estimate the minimal point. In cases where the polynomial fitting was not accurate enough, we selected the minimum value among the list of numerical results. By selecting points away from the estimated minimal point, we also checked that the uncertainty of the photon rate value due to this variation is small.  
 
For the purpose of calculating photon yield/flow, values of production rate with different parameters and for all space-time points of QGP evolution need to be obtained. The procedure of evaluating hard and soft contribution integrals then finding the minimum point is too time consuming to be performed repeatedly in order to calculate photon yield/flow. Therefore, we first obtain and tabulate photon production rates for a large set of different values for parameters $\{ \xi_1, \xi_2, \theta_q, \phi_q , q \}$. For each pair of $\{ \xi_1, \xi_2 \}$ we fit the corresponding tabulated results to the ansatz
\be 
\hat{q}\frac{dR^{\gamma}}{d^3\hat{q}} = \exp\left[-\alpha (\theta_q, \phi_q; \xi_1,\xi_2) - \beta (\theta_q, \phi_q; \xi_1,\xi_2) \hat{q} \right], \label{qansatz}
\ee
and 
\ba 
\alpha (\theta_q, \phi_q; \xi_1,\xi_2) &=&  \sum_{m=0}^{6}\sum_{n=0}^{6} a_{mn}(\xi_1, \xi_2) \Big(  {\rm sgn}(\xi_1) \cos\theta_q \Big)^{2m} \Big( {\rm sgn}(\xi_2) \cos\phi_q \Big)^{2n},  \label{alpha} \\
\beta (\theta_q, \phi_q; \xi_1,\xi_2) &=& \sum_{m=0}^{6}\sum_{n=0}^{6} b_{mn}(\xi_1, \xi_2) \Big(  {\rm sgn}(\xi_1) \cos\theta_q \Big)^{2m} \Big( {\rm sgn}(\xi_2) \cos\phi_q \Big)^{2n}, \label{beta}
\ea
where the sign function ${\rm sgn}(x)$ is used to prevent extra $\theta_q/\phi_q$ dependence of the fitted function for $\xi_1/\xi_2 = 0$ due to numerical artifacts.
We make a lookup table of coefficients $a_{mn}$ and $b_{mn}$ for each point on the grid of $\{\xi_1(i) \xi_2(j) \}$. The photon production rate for $\{\xi_1,\xi_2 \}$ values at the grid points will be calculated using \eqref{qansatz}, \eqref{alpha} and \eqref{beta}. For other values of $\{\xi_1,\xi_2 \}$ we use a linear interpolation of $\log(\hat{q}\frac{dR^{\gamma}}{d^3\hat{q}})$ values at the nearest points of the grid to $\{\xi_1,\xi_2 \}$.   

In the right panel of Fig.~\ref{plot:fig1a2} the fitting of numerically calculated photon rates to the function $\exp(-\alpha -\beta \hat{q})$ is shown for the example case of $\{\xi_1 = 9,\ \xi_2 = 2,\ \theta_q = 4\pi/10,\ \phi_q = 3\pi/10\}$ resulting in $\exp(-10.5825 - 1.44214 \ \hat{q})$ with $R_{\rm fit}^2 = 0.999951$. The uncertainty band due to variation of selected separation scale $p^*$ was not observable in the plot.  

\begin{figure}[h]
\centerline{
\includegraphics[width=1.0\linewidth]{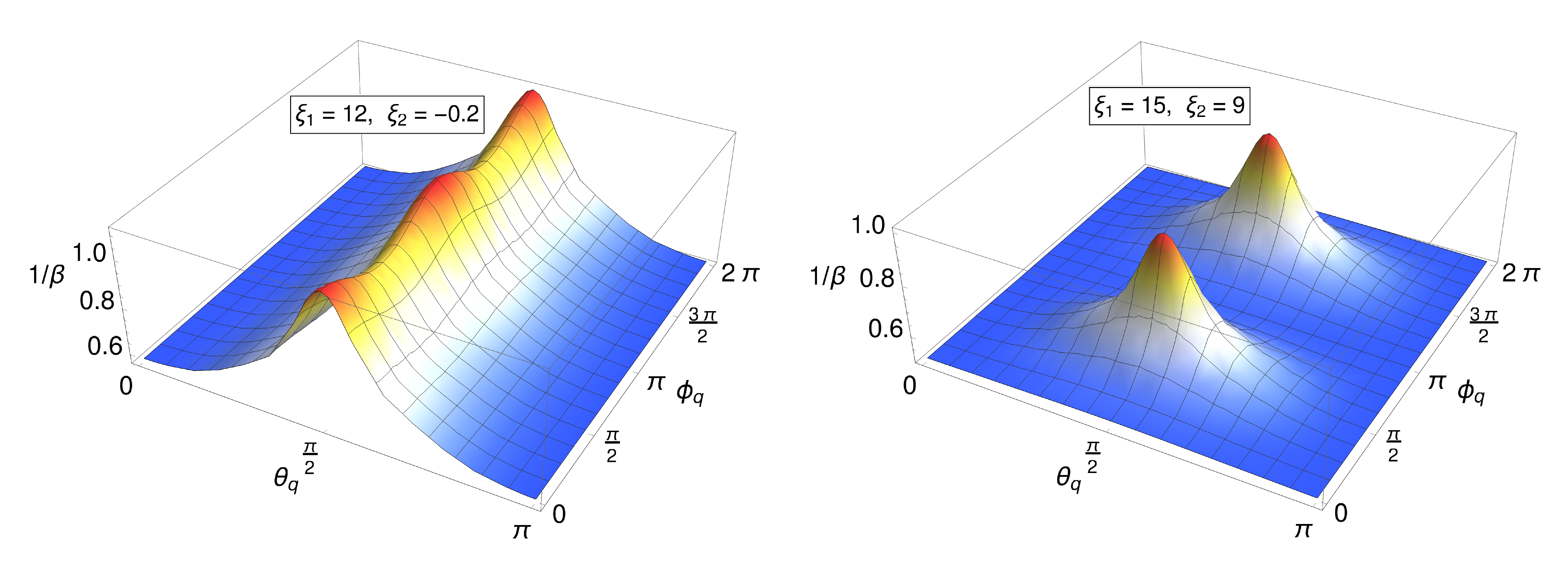}
}
\caption{Angular dependence of $1/\beta$ of ansatz \eqref{qansatz} for two example cases.} 
\label{plot:Ttf}
\end{figure}

The parameter $\beta (\theta_q, \phi_q; \xi_1,\xi_2)$ can be seen as an anisotropic rescaling factor for the inverse temperature of the radiating QGP element. One can consider $\Lambda(\xi_1,\xi_2)/\beta (\theta_q, \phi_q; \xi_1,\xi_2)$ as the anisotropic radiation temperature of QGP element in local rest frame. For isotropic QGP $\beta = 1$ and for small values of anisotropy parameters it is proportional to the original anisotropic deformation kernel $\sqrt{1+ \xi_1 \cos^2\theta + \xi_2 \sin^2\theta \cos^2\phi}$ of QGP distributions. However, in general, especially for larger anisotropy, the relation of $\alpha$ and $\beta$ to anisotropic form of QGP distributions is nontrivial and complicated.   
In Fig.~\ref{plot:Ttf} the factor $1/\beta$ as a function of $(\theta_q, \phi_q)$ is shown for $\{\xi_1 = 12,\ \xi_2 = -0.2\}$.

\section{Hydrodynamic evolution of QGP}

In order to calculate the yield and elliptic flow coefficient of real photons emitted from the QGP, we convolve the LRF anisotropic photon production rate with the space-time evolution of the strongly interacting fluid provided by 3+1d aHydro with a quasiparticle equation of state and smooth Glauber initial conditions \cite{Alqahtani:2017tnq, Alqahtani:2017jwl}. The method is basically the same as the convolution of dilepton production with aHydro evolution described in \cite{Kasmaei:2018oag}. We use the aHydro parameter values tuned to reproduce soft hadrons spectra for LHC \cite{Alqahtani:2017tnq} and RHIC \cite{Almaalol:2018gjh}. Tuned values of initial central temperature $T_0$ and shear viscosity to entropy ratio $\eta/s$ are $\{600\ {\rm MeV}, 2/4\pi \}$ for LHC and $\{455\ {\rm MeV}, 2.25/4\pi \}$ for RHIC.
In all of the results presented in this paper we only consider real photons with rapidity $y=0$ in lab frame. We only consider the photons emitted from QGP phase and we set the rate to zero for fluid elements with effective temperature $T_{\rm eff}$ below the critical temperature $T_c = 155$ MeV. 

The evolving momentum distributions in the latest version of 3+1d aHydro include a temperature dependent mass $\hat{m}$ which is calculated using isotropic lattice QCD results and matching energy densities of isotropic and anisotropic systems. To translate aHydro results to ellipsoidal distributions used for quark self-energy and photon rate calculations we neglect $\hat{m}$. This approximation, which we later check, allows one to convert  aHydro parameters $\{\alpha_{x,y,z}, \lambda \}$ to parameters $\{\xi_{1,2}, \Lambda \}$ used in photon rate by transformation
\ba
\Lambda &=& \lambda \alpha_y, \\
\xi_1 &=& \left( \frac{\alpha_y}{\alpha_z} \right)^2 -1, \\
\xi_2 &=& \left( \frac{\alpha_y}{\alpha_x} \right)^2 -1,
\ea
where $\lambda$ and $\Lambda$ are temperature-like scales of anisotropic distributions in the two parametrizations.

\begin{figure}[h]
\centerline{
\includegraphics[width=1.0\linewidth]{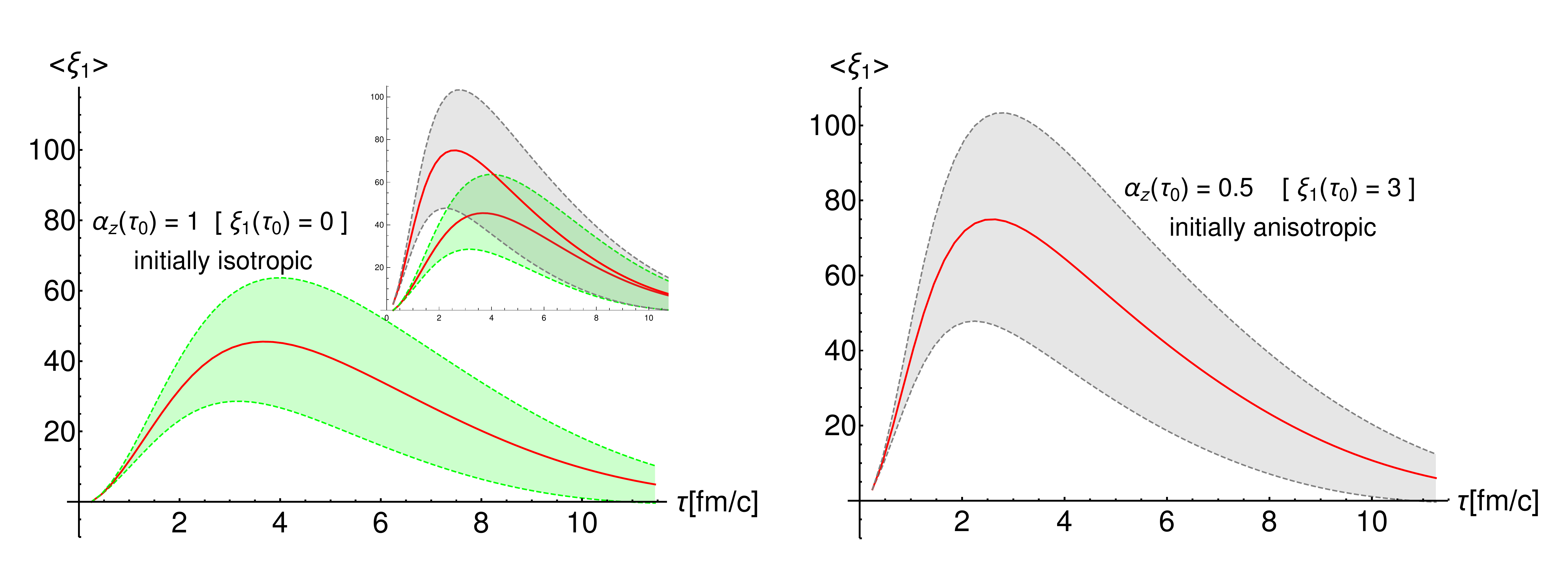}
}
\caption{Proper time evolution of spatial average of $\xi_1$ values at zero space-time rapidity hypersurface, for initially isotropic (left panel) and initially anisotropic (right panel) cases both for 30-40\% Pb-Pb collisions. Bands represent spatial standard deviations. The inset in left panel compare the two cases in one plot.} 
\label{plot:xioftau}
\end{figure}

To illustrate the evolution of momentum anisotropy, in Fig.~\ref{plot:xioftau} we plot spatial average $\langle {\xi_1} \rangle$ at zero space-time rapidity hypersurface as a function of proper time $\tau$. The bands in the plots show spatial standard deviation of $\xi_1$ values. We compare the evolution for cases with and without initial momentum anisotropy where both cases are tuned to give the best and similarly accurate fits to LHC soft hadronic spectra \cite{Kasmaei:2018oag}. The curves of $\langle {\xi_1} \rangle$ and its deviation in both cases converge at late times, but in the case with a relatively small initial anisotropy $\xi_1(\tau_0) = 3$ the values of $\xi_1$ grow to much larger values at early times. Even in the initially isotropic case, nonequilibrium dissipative effects lead to growth of momentum anisotropy before its relaxation back towards isotropy.  

\section{Results and discussion}
\label{results}

In this section we present the results for the yield and elliptic flow coefficient ($v_2$) of real photons emitted from a momentum-anisotropic QGP with space-time evolution described by aHydro. In particular we investigate the effects of initial momentum anisotropy in both longitudinal and transverse directions.

\subsection{Different centrality classes}

\begin{figure}[h]
\centerline{
\includegraphics[width=1.0\linewidth]{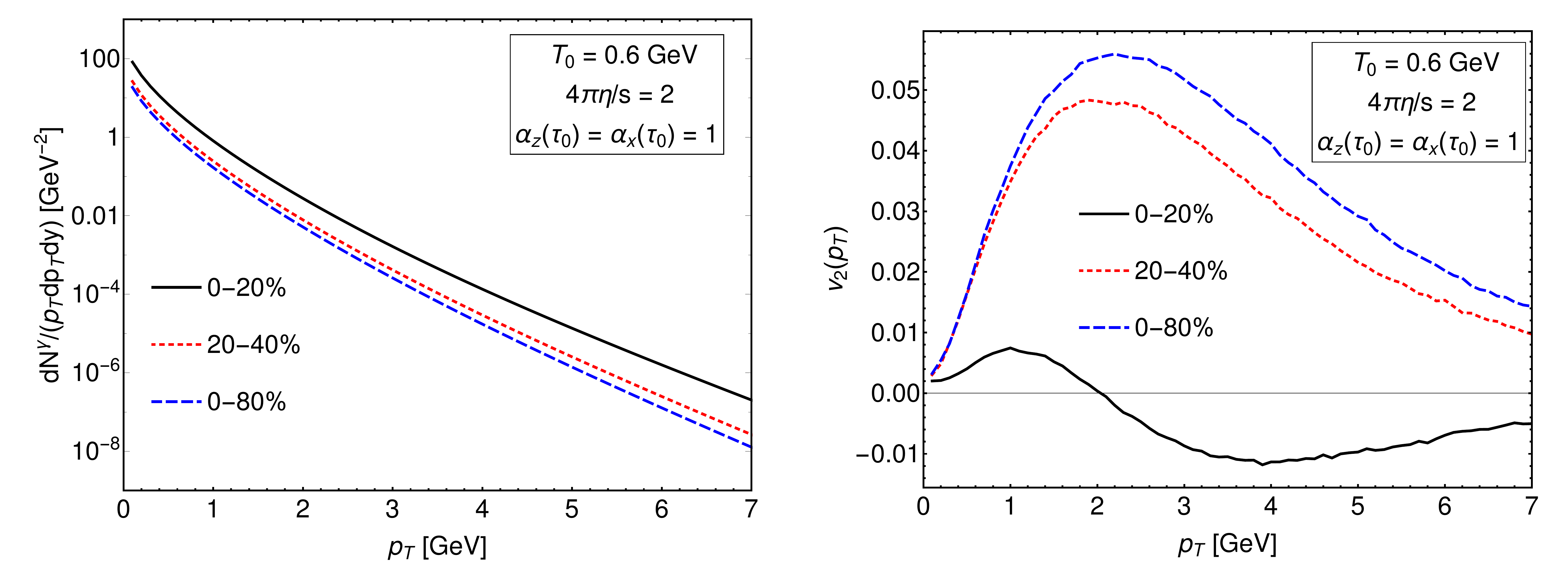}
}
\caption{Yield (left panel) and elliptic flow coefficient $v_2$ (right panel) of zero rapidity photons emitted from QGP generated at Pb-Pb collisions at LHC energy. Results are shown for three centrality classes. Initial momentum isotropy is assumed.} 
\label{plot:cent}
\end{figure}

The results for Pb-Pb collisions at $\sqrt{s}=2.76$ TeV, considering momentum-isotropic initial condition, are shown in Fig.~\ref{plot:cent} for 0-20\%, 20-40\% and 0-80\% centrality classes. The QGP photon yield in central collisions is several times higher than in peripheral ones, but their $v_2$ shows a change in sign and is negative for $p_T > 2$ GeV. The same behavior at central collisions was found for dilepton $v_2$ in our previous paper \cite{Kasmaei:2018oag}.

\subsection{Varying $\alpha_z(\tau_0)$} \label{subsec_varz}

\begin{figure}[h]
\centerline{
\includegraphics[width=1.0\linewidth]{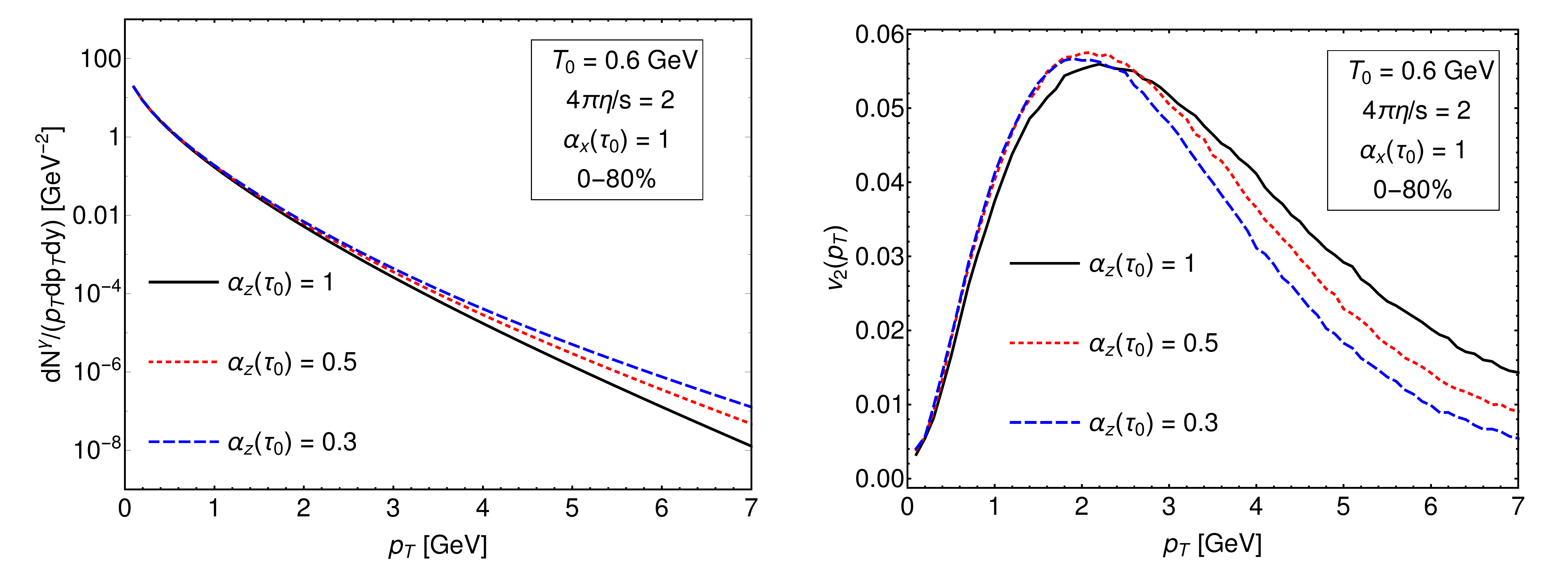}
}
\caption{Yield (left) and $v_2$ (right) of QGP photons at 0-80\% Pb-Pb collisions for different initial longitudinal momentum anisotrpies. Initial transverse momentum is assumed to be isotropic.} 
\label{plot:varaz}
\end{figure}

To investigate the effects of initial momentum anisotropy of QGP we vary the value of $\alpha_z(\tau_0)$ from $1$ (isotropic case) keeping the initial transverse momentum distribution isotropic ($\alpha_x(\tau_0) = 1$). For 0-80\% centrality class the photon yield and $v_2$ results with $\alpha_z(\tau_0) \in \{1, 0.5, 0.3 \}$ are shown in Fig.~\ref{plot:varaz}. It is observed that the results for $p_T \gtrsim 2$ GeV are sensitive to initial momentum anisotropy. The yield is increased and $v_2$ reduces for more anisotropic momentum distributions of initial conditions (lower $\alpha_z(\tau_0)$). It should be noted that with variation of initial momentum anisotropy one in general needs to retune the parameters of the model such as initial temperature to find the optimal agreement with the hadronic data. In this paper the goal is not to provide phenomenological parameter tuning and we used same $\{T_0, \eta/s \}$ for cases with different initial momentum anisotropy. In a previous paper \cite{Kasmaei:2018oag} we showed that retuning $T_0$ for an initially anisotropic QGP does not change the interpretation of the results on electromagnetic emission.  

\subsection{Varying $\alpha_x(\tau_0)$}

\begin{figure}[h]
\centerline{
\includegraphics[width=1.0\linewidth]{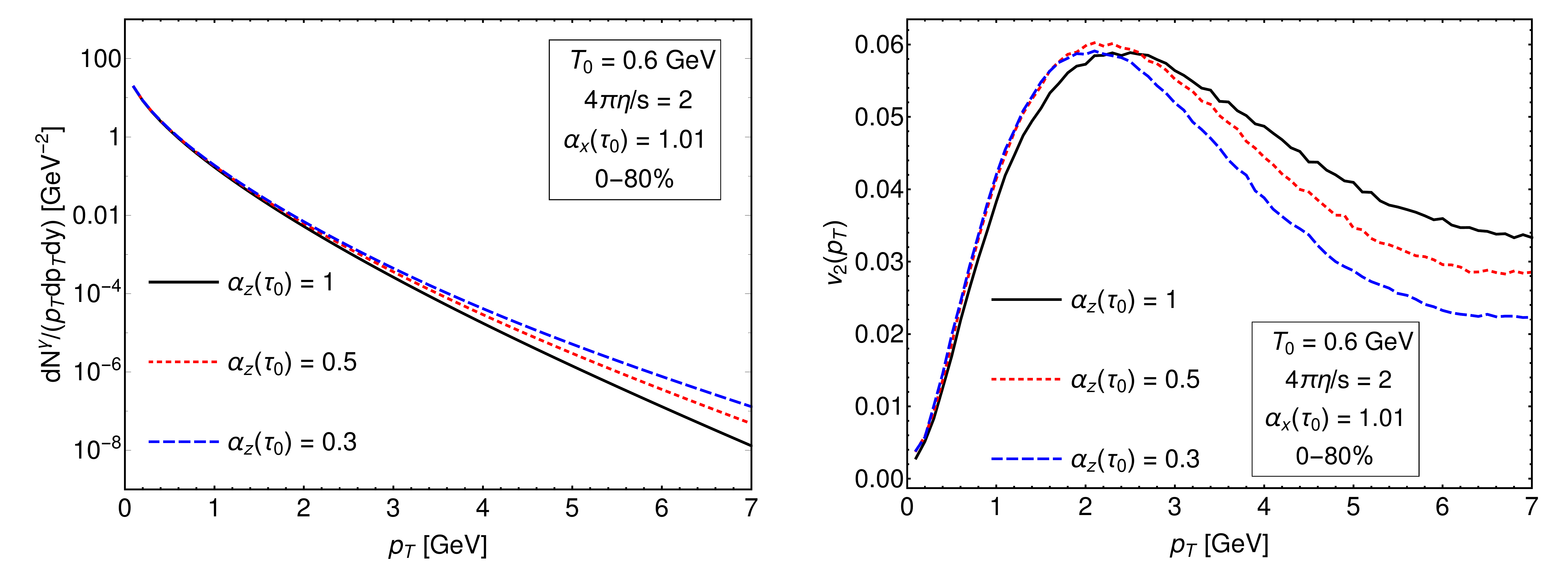}
}
\caption{Same as Fig.~\ref{plot:varaz} but assuming a small initial transverse anisotropy $\alpha_x(\tau_0) = 1.01$ of momentum distribution.} 
\label{plot:iniax}
\end{figure}

\begin{figure}[h]
\centerline{
\includegraphics[width=1.0\linewidth]{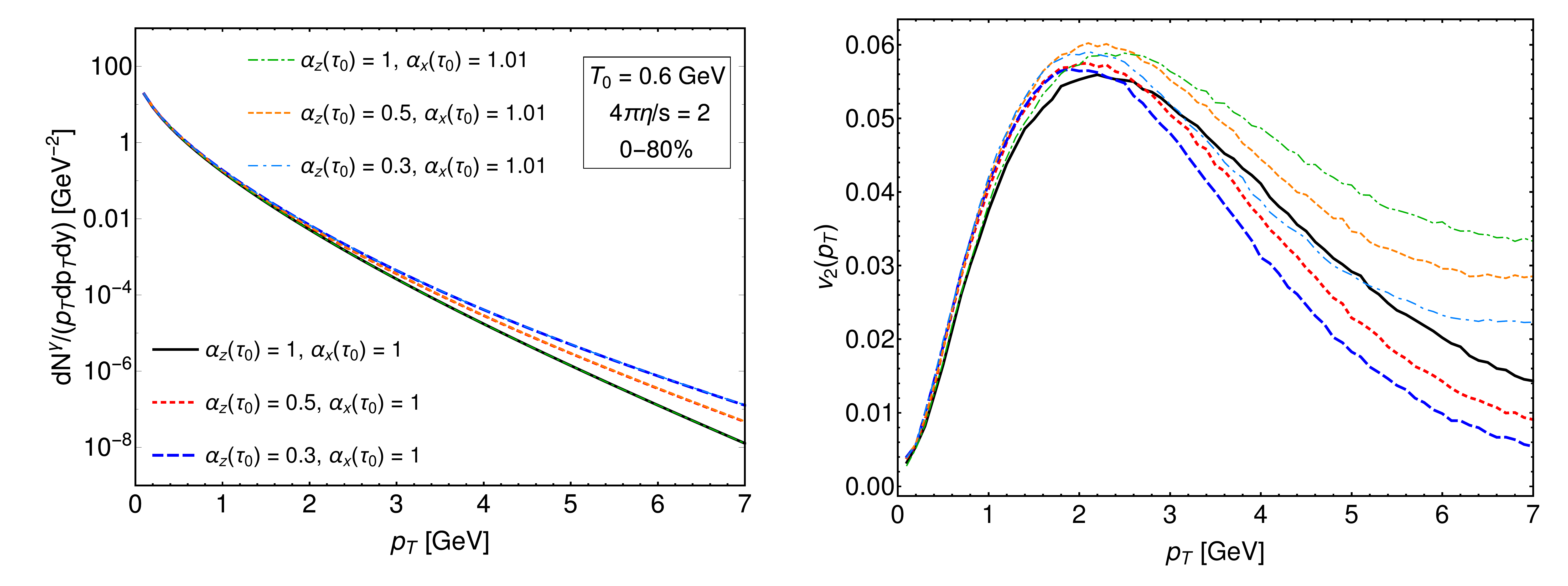}
}
\caption{The plots in Fig.~\ref{plot:varaz} and Fig.~\ref{plot:iniax} compared in same figure.} 
\label{plot:iniaxcomp}
\end{figure}

One can expect that LRF transverse momentum anisotropy can have observable effects on flow coefficients. Here we repeat the same yield and $v_2$ calculations as in Sec.~\ref{subsec_varz}, but setting $\alpha_x(\tau_0) = 1.01$. The results in Fig.~\ref{plot:iniax} and \ref{plot:iniaxcomp} show that for even a small transverse initial momentum anisotropy, QGP-emitted photon $v_2$ at $p_T \gtrsim 2$ GeV can grow to more than two times higher values while the yield is essentially unchanged. We must note that in quantitative modeling of heavy-ion collisions the proper way would be to consider the strength and direction of initial transverse momentum anisotropy as fluctuating variables event-by-event. Therefore, the large sensitivity of photon $v_2$ to $\alpha_x(\tau_0)$ as seen in the results of this paper must be considered as an overestimate and a motivation for future investigations including more relevant effects.

%
%

\subsection{Conformal approximation uncertainty}

\begin{figure}[h]
\centerline{
\includegraphics[width=1.0\linewidth]{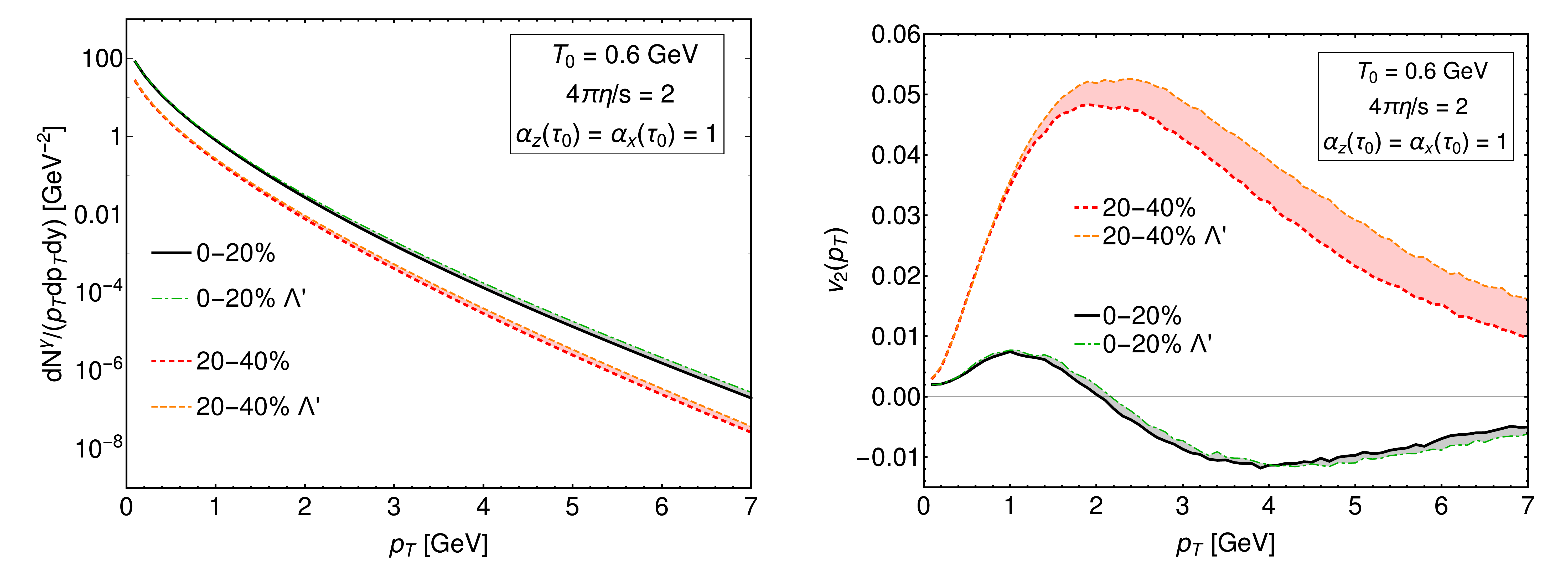}
}
\caption{Uncertainty bands due to conformal approximation neglecting $\hat{m}$ when connecting aHydro output with the LRF photon rate. Results are shown for initial momentum isotropic conditions (Pb-Pb).} 
\label{plot:lambdacent}
\end{figure}

\begin{figure}[h]
\centerline{
\includegraphics[width=1.0\linewidth]{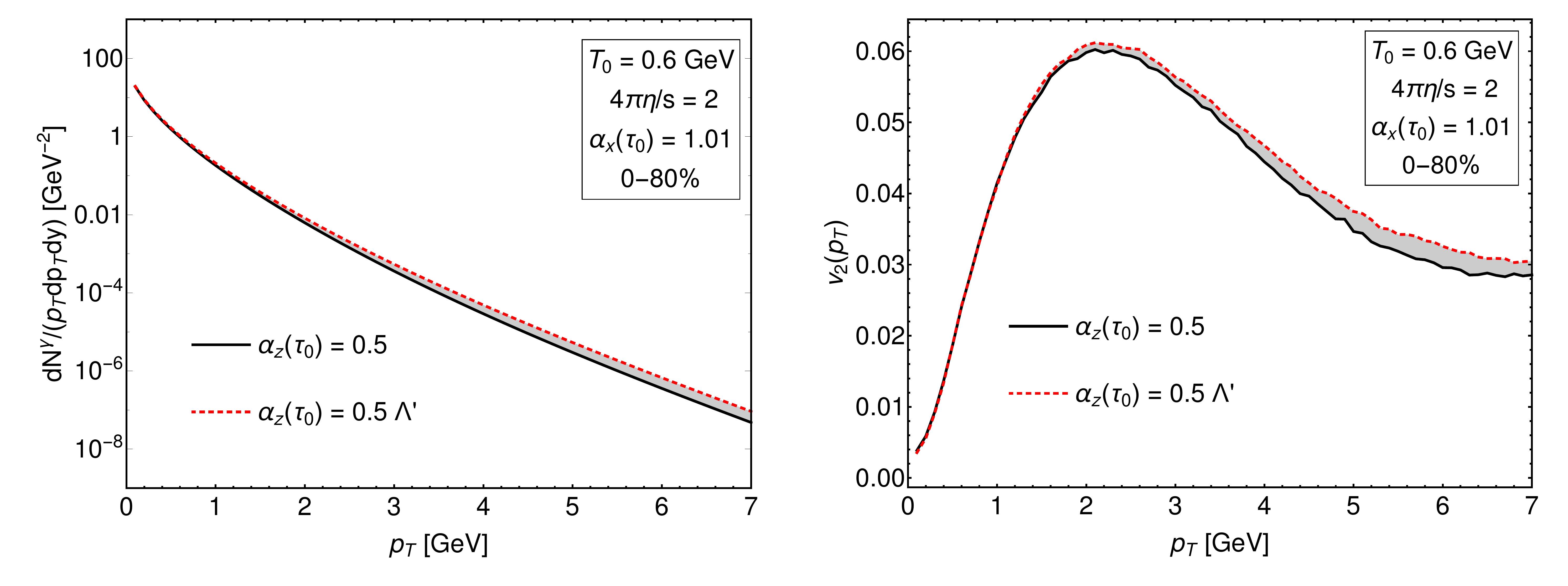}
}
\caption{Uncertainty bands due to conformal approximation for a case with initial momentum anisotropy (Pb-Pb).} 
\label{plot:lambdainiAx}
\end{figure}

In the calculations of photon yield and $v_2$ in this paper, when we convolve photon rate with aHydro results, we directly convert aHydro parameters $\{ \alpha_{x,y,z}, \lambda \}$ to LRF photon rate parameters $\{ \xi_{1,2}, \Lambda \}$ neglecting the temperature dependent mass $\hat{m}$ which is present in aHydro evolution. In order to estimate the uncertainty due to this approximation we rematch aHydro effective temperature $T_{\rm eff}(\lambda(\alpha_{x,y,z}, \hat{m}))$ to the scale $\Lambda'(\xi_{1,2})$ defined by
\be  
\left(\frac{T_{\rm eff}}{\Lambda'}\right)^4 = \frac{1}{4\pi} \int_0^{2\pi}d\phi \int_{-1}^1 \frac{d(\cos\theta)}{(1+ \xi_1 \cos^2\theta + \xi_2 \sin^2\theta \cos^2\phi)^2},
\ee 
which is basically the matching of energy densities i.e. $\epsilon(\Lambda'(\xi_{1,2})) = \epsilon_{\rm iso}(T_{\rm eff})$. We repeat the calculations of photon yield and flow using $\Lambda'(\xi_{1,2})$ instead of $\lambda(\alpha_{x,y,z}, \hat{m})$ and compare the results. The same calculations of 0-20\% and 20-40\% centrality class presented in Fig.~\ref{plot:cent} are repeated with $\Lambda'$ and compared in Fig.~\ref{plot:lambdacent}. The bands show the level of uncertainty due to conformal approximation. The effects of $\hat{m}$ on yields and photon flow of central collisions seem to be negligible and there is a notable uncertainty band for flow in 20-40\% case. However this uncertainty does not seem to change the overall estimates and qualitative interpretations of the results. In Fig.~\ref{plot:lambdainiAx} we show the uncertainty band due to conformal approximation for the case with initial momentum anisotropy $\{\alpha_z(\tau_0) = 0.5, \ \alpha_x(\tau_0) = 1.01 \}$ which seems to be small for both the photon yield and $v_2$ in 0-80\% centrality class.

\subsection{Au-Au collisions at $\sqrt{s} = $ 200 GeV}

\begin{figure}[h]
\centerline{
\includegraphics[width=1.0\linewidth]{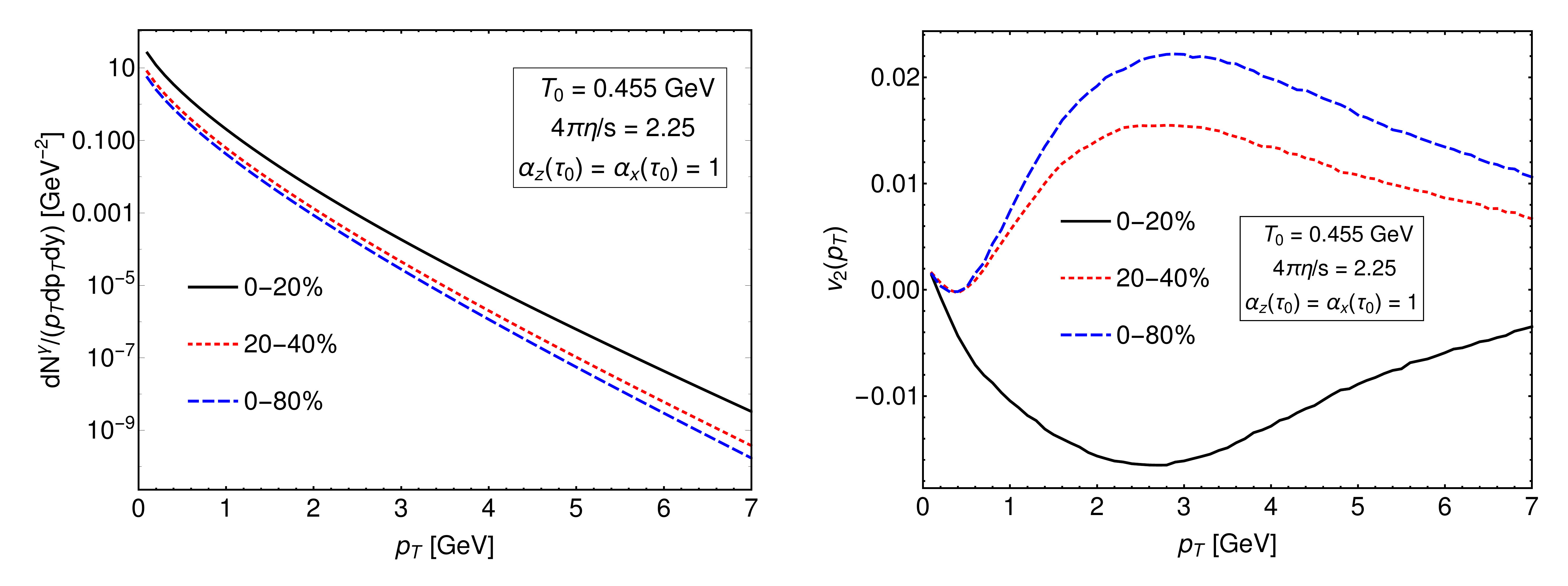}
}
\caption{Same calculations as Fig.~\ref{plot:cent} repeated for Au-Au collisions at RHIC at $\sqrt{s} =$ 200 GeV.} 
\label{plot:rhic_cent}
\end{figure}

In all of the results presented in previous sections we used aHydro model for Pb-Pb collisions at LHC at $\sqrt{s} =$ 2.76 TeV. In this section we show the results for the aHydro calculation of the yield and flow of photons emitted from QGP generated at Au-Au collisions at RHIC at $\sqrt{s} =$ 200 GeV. We use aHydro parameter values reported in \cite{Almaalol:2018gjh} by tuning to the soft hadron spectra. Photon yield and $v_2$ results for three centrality classes and momentum isotropic initial condition are shown in Fig.~\ref{plot:rhic_cent}. Comparing with similar results for Pb-Pb collisions in Fig.~\ref{plot:cent}, it can be seen that, as expected, both yield and $v_2$ for Au-Au collision are smaller than in the Pb-Pb case due to differences in $T_0$, $\eta/s$. 

\section{Conclusions and outlook}

In this paper we calculated the real photon production in a momentum anisotropic QGP. For the first time we included transverse momentum anisotropy using an ellipsoidal parametrization of the QGP parton distribution functions. This is an extension of previous results of photon production including only longitudinal momentum anisotropy \cite{Schenke:2006yp} and in line with our previous study on dilepton emission \cite{Kasmaei:2018oag}. To calculate soft contributions to the photon rate we utilized our previous results for quark self-energy in anisotropic QGP \cite{Kasmaei:2018yrr} advancing in direction of including more information about collective excitations of the strongly interacting matter in phenomenological studies. In addition, the efficient method introduced in \cite{Kasmaei:2018yrr} for self-energy calculations in anisotropic medium was essential in meeting the challenge of demanding numerical calculation of the photon rate with two extra parameters $\xi_2$ and $\phi_q$ compared to the previous spheroidally anisotropic case. We showed that the numerically calculated nonequilibrium photon rate can be accurately fitted to a function similar to the corresponding equilibrium rate by introducing direction dependence to its parameters. In particular we introduced anisotropic radiation temperature and radiation intensity for the QGP. We emphasize that these functions in general are not proportional to the anisotropic deformation introduced to the momentum distributions of QGP degrees of freedom. This idea provides us with a compact formulation of the photon rate which highly facilitates further calculations of yield and flow. We expect that similar ideas can be suitably applied to other observables such as dilepton production and modification of hard probes in nonequilibrium QGP. 

We convolved the LRF aniotropic photon rate with space-time evolution of QGP modeled by aHydro with a quasiparticle EOS to obtain the yield and elliptic flow coefficient of zero rapidity real photons generated from the QGP in Pb-Pb collisions at LHC at $\sqrt{s} =$ 2.76 TeV and Au-Au collision at RHIC at $\sqrt{s} =$ 200 GeV. With initially isotropic momentum distributions, we presented the results for 0-20\%, 20-40\% and 0-80\% centrality classes. We observed that for central collisions the anisotropic model predicts negative $v_2$ for QGP generated photons. For Pb-Pb collisions we varied the initial momentum anisotropy and showed that for $p_T \gtrsim$ 2 GeV the results are sensitive to initial longitudinal momentum anisotropy. We found a strong sensitivity of $v_2$ at $p_T \gtrsim$ 2 GeV  to the initial transverse momentum anisotropy. However, lacking a full analysis including event-by-event fluctuations and prompt photon sources we expect that our results represent an overestimate of the effects of transverse initial momentum anisotropy. We also checked the uncertainty due to conformal approximation and neglecting the thermal quasiparticle mass in our calculations of LRF rate. We showed that qualitative interpretations are not changed due to this uncertainty and overall estimates of the quantitative results can be considered reliable at this level. 

In future studies, including a more complete set of photon sources in heavy-ion collisions such as prompt and hadronic gas and decay photons can provide a more definitive phenomenological understanding and better connection to experimental results. We expect that same ideas introduced in this paper can be useful for making further progress in incorporation of various contributions such as nonequilibrium, nonperturbative and magnetic field effects in photon production rate calculation.

\section*{Acknowledgement}

B. S. Kasmaei and M. Strickland were supported by
the U.S. Department of Energy, Office of Science, Office of Nuclear Physics under Award No. DE-SC0013470.

\bibliography{photon_paper}

\end{document}